# Learning Soil Physics from Partial Knowledge and Data: Partitioning Capillary and Adsorbed Soil Water

Sarem Norouzi[1,3], Per Moldrup[2], Ben Moseley[3], David Robinson[4], Dani Or[5], Tobias L. Hohenbrink[6], Budiman Minasny[7], Morteza Sadeghi[8], Emmanuel Arthur[1], Markus Tuller[9†], Mogens H. Greve[1], and Lis W. de Jonge[1]

[1] Department of Agroecology, Aarhus University, Tjele, Denmark

[2] Department of the Built Environment, Aalborg University, Aalborg, Denmark

[3] Department of Earth Sciences and Engineering, Imperial College London, London, UK

[4] UK Centre for Ecology & Hydrology, Bangor, UK

[5] Department of Civil and Environmental Engineering, University of Nevada, Reno, NV, USA

[6] German Weather Service (DWD), Agrometeorological Research Center, Braunschweig, Germany

[7] School of Life and Environmental Sciences, The University of Sydney, Australia

[8] California Department of Water Resources, Sacramento, USA

[9] Department of Environmental Science, The University of Arizona, Tucson, AZ, USA

Corresponding author: Sarem Norouzi (sarem.nrz@agro.au.dk)

[†] The late Markus Tuller contributed substantially to this research; he passed away before the manuscript was submitted.

**Key Points**

- A differentiable hybrid modeling (DHM) framework integrates physics and data.
- The model partitions soil water retention into capillary and adsorbed components.
- Capillary–adsorbed transition is learned from data without pore-scale assumptions.
- DHM enables physics-informed, data-driven discovery of fundamental soil processes.

**Abstract**

Soil physics models have long relied on simplifying assumptions to represent complex processes, yet such assumptions can strongly bias model predictions. Here, we propose a paradigm-shifting differentiable hybrid modeling (DHM) framework that instead of simplifying the unknown, learns it from data. As a proof of concept, we apply the hybrid approach to the challenge of partitioning the soil water retention curve (SWRC) into capillary and adsorbed water components, a problem where traditional assumptions have led to divergent results. The hybrid framework derives this partitioning directly from data while remaining guided by a few parsimonious and universally accepted physical constraints. Using basic soil physical properties as inputs, the hybrid model couples an analytical formula for the dry end of the SWRC with data-driven physics-informed neural networks that learn the wet end, the transition between the two ends, and key soil-specific parameters. The model was trained on a SWRC dataset from 482 undisturbed soil samples from Central Europe, spanning a broad range of soil texture classes and organic carbon contents. The hybrid model successfully learned both the overall shape and the capillary and adsorbed components of the SWRC. Notably, the model revealed physically meaningful pore-scale features without relying on explicit geometrical assumptions about soil pore shape or its distribution. Moreover, the model revealed a distinctly nonlinear transition between capillary and adsorbed domains, challenging the linear assumptions invoked in previous studies. The methodology introduced here provides a blueprint for learning other soil processes where high-quality datasets are available but mechanistic understanding is incomplete.

**Keywords:** *Scientific machine learning, Differentiable modeling, Soil hydraulic properties, Hybrid modeling*

# 1. Introduction

Physics-based modeling approaches have a long tradition in soil physics and have been applied to simulate fundamental soil processes such as infiltration, evaporation, solute transport, and energy exchange in the vadose zone (Green and Ampt, 1911; Gardner, 1958; Philip and De Vries, 1957; Van Genuchten, 1982). Mechanistic modeling typically involves conceptualizing the problem, deriving governing equations from physical laws or empirical relationships, and validating the resulting models against experimental data. These models are then used to study system functions and behaviors, test hypotheses, and assess the responses of a system to changes in the driving forces or internal properties.

The process of deriving representative models in any natural system inevitably requires simplifying poorly understood components of the system. For instance, soil physics makes assumptions regarding (largely unobservable) pore geometry and its distribution within soil, attainment of equilibrium conditions, or the functional forms of constitutive relationships. While these simplifications enable tractable formulations, they are shaped by the modeler's view of the system that may bias the true representation of soil processes. While some assumptions are refined as new evidence emerges, the original modeling and representation bias may persist.

To overcome some of these challenges, we propose differentiable hybrid modeling (DHM) as an alternative approach to the explicit representation of certain soil physical processes (e.g., making assumptions about pore sizes, shapes and their distributions). Hybrid methods, which fall under the broader umbrella of scientific machine learning (SciML), embed neural networks within physical models so that the unknown or poorly understood components of a system can be learned directly from data while the well-established physical laws remain explicitly enforced (Psichogios and Ungar, 1992; Karniadakis et al., 2021; Moseley, 2022; Shen et al., 2023). By constructing both the physical equations and the neural components in a differentiable form, these hybrid systems can be trained end-to-end using gradient-based optimization. This allows all parameters to be adjusted jointly to minimize a downstream, physics-informed loss function. This dual nature of these methods maintains the interpretability of traditional formulations while enabling discovery of processes that are otherwise inaccessible through purely mechanistic or purely empirical approaches.

Automatic differentiation (AD) which is the backbone of DHM (Baydin et al., 2018), has also advanced other domains of SciML such as physics-informed neural networks (PINNs) (Raissi et al., 2019). PINNs have been successfully applied in vadose zone modeling to estimate soil hydraulic properties from soil moisture measurements (Tartakovsky et al., 2020; Bandai and Ghezzehei, 2021; Minasny et al., 2024), to model water flow and solute transport using geoelectrical data (Haruzi and Moreno, 2023), and to develop flexible non-parametric pedotransfer functions (PTFs) for the soil water retention curve (Norouzi et al., 2025). While DHM shares conceptual similarities with PINNs, it differs fundamentally in structure and training objective. PINNs learn solutions that satisfy governing physical laws, while DHM embeds neural networks within an existing analytical model or process-based model to learn the unknown or less understood components.

To demonstrate how DHM can be used as a framework for learning complex soil physical processes, we apply this method to the classic problem of modeling the soil water retention curve and its partitioning to capillary and adsorptive components. The SWRC is a fundamental soil property that defines the relationship between water content and matric head (or suction) in soil. Modeling a variety of land-surface processes, including infiltration, runoff, evaporation, and energy exchange at the land surface requires knowledge of the SWRC across scales (Gupta et al., 2022; Tehrani et al., 2023; Turek et al., 2025). Moreover, partitioning the SWRC into its capillary and adsorbed components enhances the modeling of the soil hydraulic conductivity curve and is essential for determining the liquid–water interfacial area, which influences soil health and biogeochemical processes, particularly the retention and transport of interfacially active contaminants such as PFAS (Guo et al., 2020; Brusseau, 2023).

All existing SWRC models that partition capillary and adsorbed film water have been developed based on specific prior assumptions and simplifications about soil pore geometry, the functional forms used to describe water retention components, and the transition between these two soil water regimes. Because these assumptions directly influence the resulting partitioning, different models applied to the same soil can yield substantially different outcomes, reflecting the sensitivity of predictions to their underlying assumptions (Or and Tuller, 1999; Lebeau and Konrad, 2010; Peters, 2013; Lu, 2016; Weber et al., 2019; Ghorbani et al., 2025). Therefore, there remains a need for new approaches that can learn the partitioning directly from data with minimal prior assumptions, while still respecting key physical constraints.

In this study, we develop a hybrid model that learns the shape of the SWRC as well as the capillary and adsorbed water content components from basic soil properties using a state-of-the-art differentiable modeling approach. In key contrast to traditional parametric models with rigid physical assumptions, our method relaxes these assumptions and instead learns a flexible, physically interpretable, and data-driven partitioning, with only universally accepted assumptions included. Our study introduces a new generation of SWRC models, which we term "semi-parametric" (i.e., semi-analytical) models. In this new type of SWRC models, part of the curve is described by analytical equations, while the remaining parts are flexibly learned from data using neural networks, yet the final SWRC remains continuous, differentiable, and physically consistent.

We believe our work demonstrates the potential of the DHM framework for unifying physical theory and data-driven discovery across a wide range of fundamental soil physical processes.

## 2. Materials and Methods

### *2.1. Conventional Modeling Approach*

Classical models of soil water retention partition the total water content into capillary and adsorbed (film) domains by assuming an idealized pore geometry and prescribed pore-size distributions. For example, Or and Tuller (1999) represented the pore space as a unit cell consisting of an angular central pore (e.g., square or polygonal) connected to slit-shaped corners that provide surface area for film adsorption. By applying the augmented Young–Laplace equation, they calculated liquid distribution within this unit cell as a function of both capillary and adsorptive contributions. The model was then upscaled to the sample scale through statistical averaging over an assumed pore-size distribution.

In semi-empirical approaches, the capillary domain is modeled by representing soil pores as a bundle of cylindrical tubes that conform to a specified size distribution (e.g., lognormal). The adsorbed water component is described empirically, and the transition between capillary and film domains is represented by a simple linear or additive formulation.

Different choices of pore geometry, pore-size distribution, and functional relationships have led to a variety of models in the literature that often produce divergent capillary–adsorbed partitioning when applied to the same soil.

## 2.2. Hybrid Modeling Framework

Our hybrid modeling approach aims to construct a data-driven framework that flexibly learns both the overall shape of the SWRC and the individual contributions of capillary and adsorbed water from basic soil properties. Specifically, we aim to relax the restrictive assumptions in conventional models: idealized pore shapes, certain pore size distributions, fixed functional forms for which there is no universal agreement or any assumption about the transition between the two components of soil water.

The hybrid model uses the same inputs and outputs as conventional pedotransfer functions, translating basic soil physical properties to SWRC. However, during this mapping from inputs to outputs, the model learns multiple intermediate processes implicitly, without requiring explicit data for them.

We begin with universally accepted principles describing the coexistence and behavior of capillary and adsorptive forces across the full moisture range. As we derive the governing equations from these known principles, any unknown terms are treated as components to be inferred from data. In this strategy, the semi-analytical expression provides the physical skeleton of the model while the unknown parts are flexibly learned from data. To physically guide the model, we only rely on a few physical constraints that are universally accepted.

### 2.2.1. Physical definitions and model derivation

Capillary water refers to liquid water filling the spaces between soil particles, held by surface tension and the contact angle of water with solid surfaces, which leads to the formation of curved liquid–vapor interfaces (menisci). The adsorbed film water component refers specifically to liquid water retained in thin films by adsorptive forces, where a distinct liquid–air interface is present (see for example figure 4 of Nachum (2025)). The adsorptive forces in soil arise from intermolecular interactions between the liquid and solid surfaces, including van der Waals forces, electrostatic double-layer forces, and structural (hydration) forces (Derjaguin et al., 1987; Tuller and Or, 2005a).

The total volumetric water content retained in a soil can thus be expressed as the sum of its capillary ($\theta_c$) and adsorbed ($\theta_a$) components:

$$\theta = \theta_c + \theta_a \tag{1}$$

Note that all terms in Eq. (1) are functions of matric head. As a soil dries, capillary water drains from larger pores and recedes into pore corners. Beyond a certain matric head threshold, water persists primarily as thin films adsorbed onto particle surfaces. At this dry end, experimental studies show that the water retention curve becomes linear in $pF - \theta$ space, where $pF = \log|h|$ and $h$ is the matric head in cm. This linear behavior can be described analytically by the Campbell and Shiozawa (1992) model (hereinafter referred to as Campbell-Shiozawa model and denoted by the subscript CS), which in $pF - \theta$ space is written as:

$$\theta_{CS} = \left(1 - \frac{pF}{pF_{dry}}\right)\theta_o \tag{2}$$

Where $\theta_{CS}$ is the predicted water content by Campbell-Shiozawa model, and $\theta_o$ and $pF_{dry}$ are its fitting parameters that determine the slope and intercept of this empirical model at the dry end. The parameter $pF_{dry}$ corresponds to the logarithm of matric head at oven dryness, where the soil is assumed to reach zero water content.

The Campbell–Shiozawa model was originally developed for the dry end of the SWRC (i.e., the higher range of $pF$ values), where only adsorptive forces are active. In the lower range of $pF$ values, where capillary water begins to contribute, the expression for $\theta_{cs}$ no longer holds and must be revised. To account for this, we introduce a transition function, denoted as $f$, that modifies $\theta_{cs}$ in the mixed region, where capillary and adsorbed water coexist. This function is treated as an unknown to be learned from data, and it is expressed as a function of capillary saturation, defined as $S_c = \theta_c/\theta_s$, where $\theta_s$ is the saturated water content. Accordingly, the adsorbed film component of the SWRC can be modeled as:

$$\theta_a = f(S_c)\,\theta_{cs} \tag{3}$$

Combining Eqs. (3) and (1), the total water can be expressed as:

$$\theta = \theta_c + \overbrace{f(S_c)\theta_{cs}}^{\theta_a} \tag{4}$$

The second term on the right-hand side of Eq. (4) accounts for the adsorbed film water contribution. When $\theta_c$ approaches zero (i.e., at very high $pF$ values), the retention behavior is dominated by adsorbed water. In this limit, Eq. (4) should reduce to the Campbell-Shiozawa model, Eq. (2). To ensure this, we reparametrize the transition function with a hard constraint that enforces $f(0) = 1$:

$$f(S_c) = 1 + S_c \cdot g(S_c) \tag{5}$$

where $g(S_c)$ is an unknown function that is learned from data. Combining this equation with Eq. (4), the total water content can be expressed as:

$$\theta = \theta_c + \overbrace{[1 + S_c\,g(S_c)]\theta_{cs}}^{\theta_a} \tag{6}$$

Inserting Eq. (2) into Eq. (6), we obtain:

$$\theta = \theta_c + [1 + S_c \; g(S_c)]\left(1 - \frac{pF}{pF_{dry}}\right)\theta_o \tag{7}$$

It is worth noting that Eq. (7) reflects only a general structural formulation based on a few widely accepted assumptions about water retention in soils. Specifically, it assumes that total water content consists of two components, capillary and adsorbed water, and that the contribution of the capillary component vanishes as capillary saturation ($S_c$) approaches zero. In this condition, Eq. (7) reduces to Campbell-Shiozawa model for the dry end. The function $g(S_c)$ captures the transition between capillary and adsorbed dominant regions behavior without explicitly specifying its shape and form in advance. Similarly, the capillary component, $\theta_c$, as well as soil constants (i.e., $\theta_s$, $pF_{dry}$, and $\theta_o$) are treated as unknowns to be learned from data.

### 2.2.2. Neural sub-model for the capillary water

We replace the capillary water content ($\theta_c$) in Eq. (7), which is a function of $pF$, with a dedicated neural network sub-model named $NN_c(\mathbf{x}, pF; \boldsymbol{\phi}_c)$, which receives the vector of soil physical properties $\mathbf{x} = [Sand, Silt, Clay, OC, BD]$ and $pF$ as input. The vector $\boldsymbol{\phi}_c$ represents the trainable parameters (weights and biases) of this neural network. This network has two hidden layers, each with eight units (i.e., nodes) (Table 1). By taking $pF$ as an input, the network outputs the capillary water content at the specified $pF$. This architecture produces a continuous representation of the curve without limiting it to a specific functional form (Haghverdi et al., 2012; Norouzi et al., 2025). To ensure meaningful predictions, the sub-network output is constrained to remain less than saturated water content ($\theta_s$). To enforce this, the raw output of the capillary network is passed through a sigmoid activation and scaled by the $\theta_s$ value.

*2.2.3. Neural sub-model for the transition function*

The transition function in Eqs. (5) and (7) includes an unknown component, $g(S_c)$, which maps the input $S_c$ to a scalar output. Rather than assuming a fixed analytical form for $g(S_c)$, we replace it with a fully connected neural network, $NN_g(S_c; \boldsymbol{\phi}_g)$, and allow it to be learned from data. Similarly, $\boldsymbol{\phi}_g$ refers to the trainable parameters of this network. This neural network is capable of approximating a wide range of continuous functions, which helps capture complex transition behaviors. The neural network $NN_g(S_c; \boldsymbol{\phi}_g)$ consists of two hidden layers and a linear output layer without any constraint on the output value (Table 1). We designed this sub-network to be more flexible to ensure that the shape of the overall transition function, $f(S_c)$, is not limited by the capacity of $NN_g(S_c; \boldsymbol{\phi}_g)$.

*2.2.4. Neural sub-model for the soil dependent constants*

The Campbell–Shiozawa model and the transition function both depend on three key soil-specific parameters: $\theta_s$, $pF_{dry}$, and $\theta_o$. We assume that these parameters can be predicted from basic soil physical properties, and therefore, we model each parameter using a dedicated neural network. In our hybrid framework, each parameter is estimated by a separate sub-network: $NN_s(\mathbf{x}; \boldsymbol{\phi}_s)$, $NN_{dry}(\mathbf{x}; \boldsymbol{\phi}_{dry})$, and $NN_o(\mathbf{x}; \boldsymbol{\phi}_o)$, corresponding to $\theta_s$, $pF_{dry}$, and $\theta_o$, respectively. Each sub-network takes five fundamental soil properties as defined by $\mathbf{x}$: sand, silt, clay, OC, and BD. Additionally, each of the neural networks has two hidden layers, each containing four units with Exponential Linear Unit (ELU) activation function (Table 1). Although a single multi-output neural network could have been used, given that the inputs of these networks are identical, we opted for separate networks to maintain clarity and interpretability in the modeling framework.

Since $\theta_s$ and $\theta_o$ cannot exceed unity, we constrain the predictions of $NN_s(\mathbf{x}; \boldsymbol{\phi}_s)$ and $NN_o(\mathbf{x}; \boldsymbol{\phi}_o)$ to a range between 0 and 1 by applying a sigmoid activation function in the output layers of these sub-networks. Similarly, several studies showed that the range of $pF_{dry}$ falls between 6.5 and 7.45 (Schneider and Goss, 2012; Arthur et al., 2013; Lu and Khorshidi, 2015; Karup et al., 2017). To ensure broader applicability and account for potential variability beyond these observations, we constrain the $NN_{dry}(\mathbf{x}; \boldsymbol{\phi}_{dry})$ sub-model output to predict values within a slightly wider interval of 6.2 to 7.6. This is achieved by scaling the sigmoid output of $NN_{dry}(\mathbf{x}; \boldsymbol{\phi}_{dry})$ to this target range. To keep notation simple, the explicit dependence on inputs $(\mathbf{x}, pF)$ and trainable parameters $(\boldsymbol{\phi})$ is omitted in the remaining text wherever it does not cause ambiguity.

**Table 1.** Summary of the architecture and configuration of the subnetworks.

| Network | Inputs[a] | Hidden layers[b] | Output | Output activation and scaling |
|---|---|---|---|---|
| $NN_c$ | Sa, Si, Cl, OC, BD, pF | 2 × Dense (8) (ELU) | $\theta_c$ | Sigmoid × $NN_s$ |
| $NN_g$ | $NN_c$ / $NN_s$ | 2 × Dense (16) (ELU) | $g(S_c)$ | Linear |
| $NN_s$ | Sa, Si, Cl, OC, BD | 2 × Dense (4) (ELU) | $\theta_s$ | Sigmoid × $NN_s$ |
| $NN_o$ | Sa, Si, Cl, OC, BD | 2 × Dense (4) (ELU) | $\theta_o$ | Sigmoid |
| $NN_{dry}$ | Sa, Si, Cl, OC, BD | 2 × Dense (4) (ELU) | $pF_{dry}$ | 6.2 + Sigmoid × (7.6 – 6.2) |

[a] Sa, Si, Cl, OC, and BD stand for sand, silt, clay, organic carbon, and bulk density, respectively.
[b] This column represents the number of hidden layers, the layer type, the number of units per layer, and the activation function for each neural network (NN).

### 2.2.5 Hybrid model and loss function

By embedding all the neural network sub-models into Eq. (7), we obtain our hybrid model, expressed as:

$$\theta = NN_c + \overbrace{\left[1 + \left(\frac{NN_c}{NN_s}\right) NN_g\right]\left(1 - \frac{pF}{NN_{dry}}\right) NN_o}^{\theta_a} \tag{8}$$

The total water content in this model is estimated by jointly training several neural network sub-models. It is important to note that Eq. (8) is directly trained end-to-end on measurements of the total water content as a function of $pF$ (Fig. 1), and all the neural network parameters are learned jointly. Therefore, separate training of individual sub-networks using distinct target measurements is not needed. Instead, the sub-networks adjust their predictions during training so that the final predicted total water content closely matches the measured total water content at each $pF$ value (Fig. 1).

In our hybrid modeling design, each measured point on the SWRC is treated as an independent training example, paired with its corresponding soil properties: sand, silt, clay, OC, BD, and $pF$. The key assumption is that, given a sufficient number of measurements, training the hybrid model in Eq. (8) enables the model to capture the complete shape of the SWRC, including both the capillary and adsorbed water components, which are modeled with various internal sub-networks. A major advantage of this method is that it allows samples with very few measured points to be included in the training set. This capability is not achievable with conventional parametric models, where a minimum number of measured points from each sample is needed (Rasoulzadeh et al., 2025).

The loss function used for training the hybrid model is as follows:

$$J = \frac{\lambda_1}{N_{wet}} \sum_{i=1}^{N_{wet}} [\hat{\theta}^{(i)} - \theta^{(i)}]^2 + \frac{\lambda_2}{N_{dry}} \sum_{i=1}^{N_{dry}} [\hat{\theta}^{(i)} - \theta^{(i)}]^2 + \frac{\lambda_3}{S_1} \sum_{i=1}^{S_1} (\hat{\theta}_c^2)^{(i)} + \frac{\lambda_4}{S_4} \sum_{i=1}^{S_2} \left|\frac{\partial \hat{\theta}}{\partial pF}\right|^{(i)} \tag{9}$$

Where $\hat{\theta}$ and $\theta$ are the predicted and measured water contents, respectively, and they are both a function of $pF$. The first two terms on the right-hand side of Eq. (9) represent the mean squared error between the volumetric water contents predicted by the neural network and the observed measurements, where $N_{wet}$ and $N_{dry}$ denote the number of training examples from the wet and dry ends, respectively. As shown by Norouzi et al. (2025), using separate terms for the wet-end (i.e., $pF \leq 4.2$) and dry-end (i.e., $pF > 4.2$) is necessary to account for disparities in sample sizes and the narrower range of water contents typically observed at the dry end. The $\lambda$ coefficients are weights assigned to each term in the loss function which determine the relative importance of each loss component during training.

The last two terms in Eq. (9) are introduced to enforce two physics-based constraints described below.

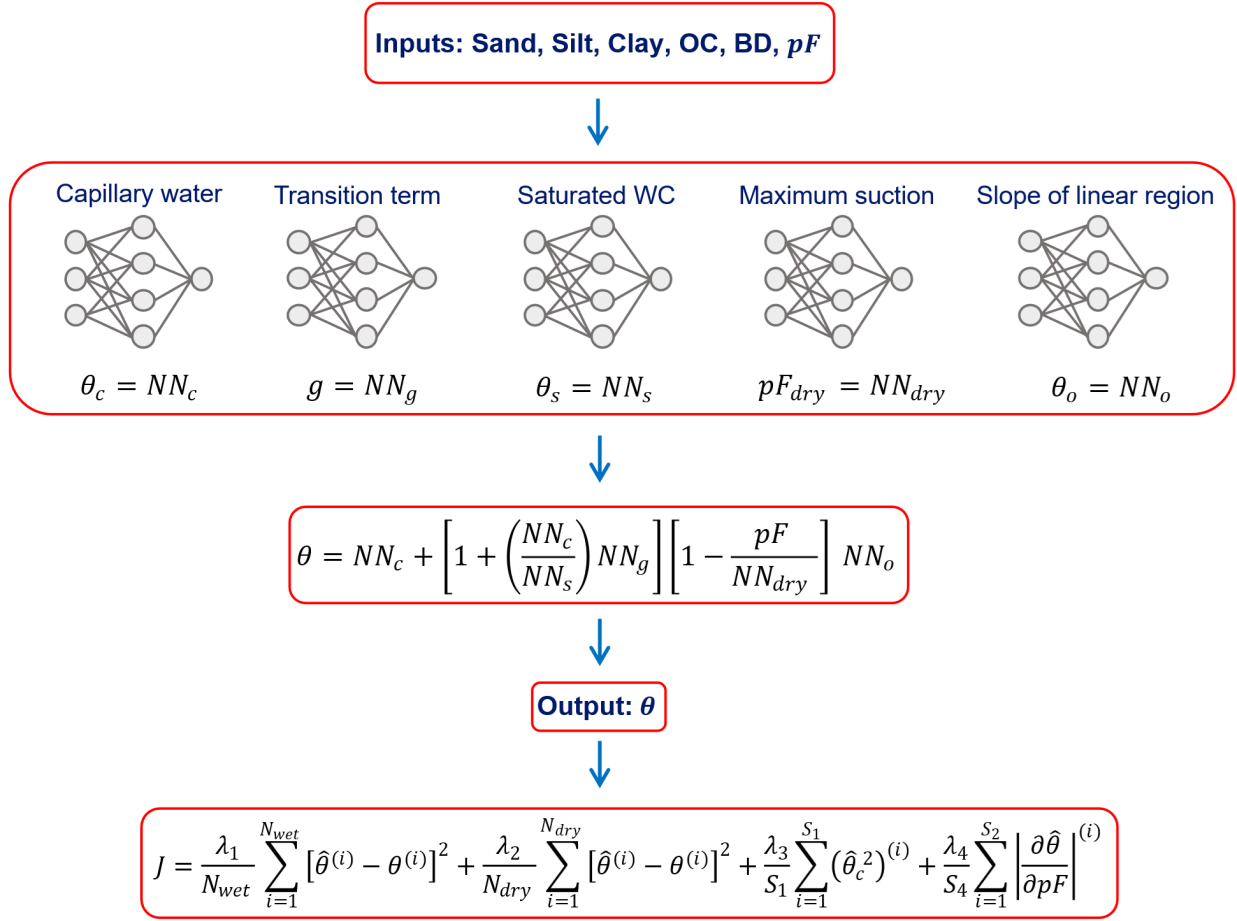

**Figure 1.** Workflow of the proposed hybrid model illustrating the interior neural networks, analytical formulation, and loss function structure for learning the soil water retention curve and its capillary and adsorbed components. The model inputs, including soil physical properties and $pF$, are fed into separate neural networks whose outputs are combined within the hybrid formulation to predict the total water content, which is then used in the loss function. The entire model, including all neural network components, is trained end-to-end, meaning that all parameters are optimized jointly from input to output through gradient-based minimization of a downstream, physics-informed loss function, without explicit labeled data for the individual sub-networks. $pF$ serves as an input only to the capillary neural network ($NN_c$).

*2.2.6 Universally accepted physical constraints*

Our hybrid model relies solely on general physical reasoning without imposing rigid or system-specific assumptions. Here, to guide the model, we incorporate two physical constraints that are broadly accepted in soil physics and supported by pore-scale understanding.

First, at high suctions in soil (i.e., for $pF > 5$), the soil water content is assumed to be entirely in adsorbed form, meaning that the capillary water content should approach zero in this range of $pF$ (Norouzi et al., 2025; Tuller and Or, 2005b). To enforce this, we introduce a constraint in the loss function. Specifically, we generate a set of residual (collocation) points, which are synthetic samples with random combinations of sand, silt, clay, OC, and BD, paired with random $pF$ values higher than 5. During each training step, the output of the $\theta_c$ neural network at these residual points is computed, and the mean of the squared values is added as a penalty term to the custom loss function to encourage $\theta_c$ to approach zero for $pF > 5$ (third term in Eq. 9). These residual points are illustrated in Fig. 2a.

Second, the soil air-entry value, also known as the bubbling pressure, corresponds to the matric head at which air begins to penetrate the largest soil pores (Fredlund and Xing, 1994; Sourmanabad et al., 2024). According to this definition, when the soil matric head (expressed in terms of $pF$) is below a specific value, the soil remains saturated, and its water content remains constant despite further changes in matric head. This condition can be mathematically represented as:

$$\frac{d\theta}{dpF} = 0, \qquad pF < pF_{air-entry} \tag{10}$$

The air-entry value, $pF_{air-entry}$, depends on both soil texture and structure and is generally lower in coarse-textured soils (Rawls et al., 1982). To implement this constraint within our neural network, we set $pF = -0.3$ (equivalent to a matric head of -0.5 cm) as the minimum threshold, below which the soil is assumed to stay fully saturated, with water content remaining constant despite variations in $pF$. As illustrated in Fig. 2b, we generate a set of residual points, which are randomly sampled within the input space using $pF$ values from -0.3 and -2.0. During training, we evaluate Eq. (10) at these points and penalize deviations from this constraint by adding the fourth term to the loss function (Eq. 9).

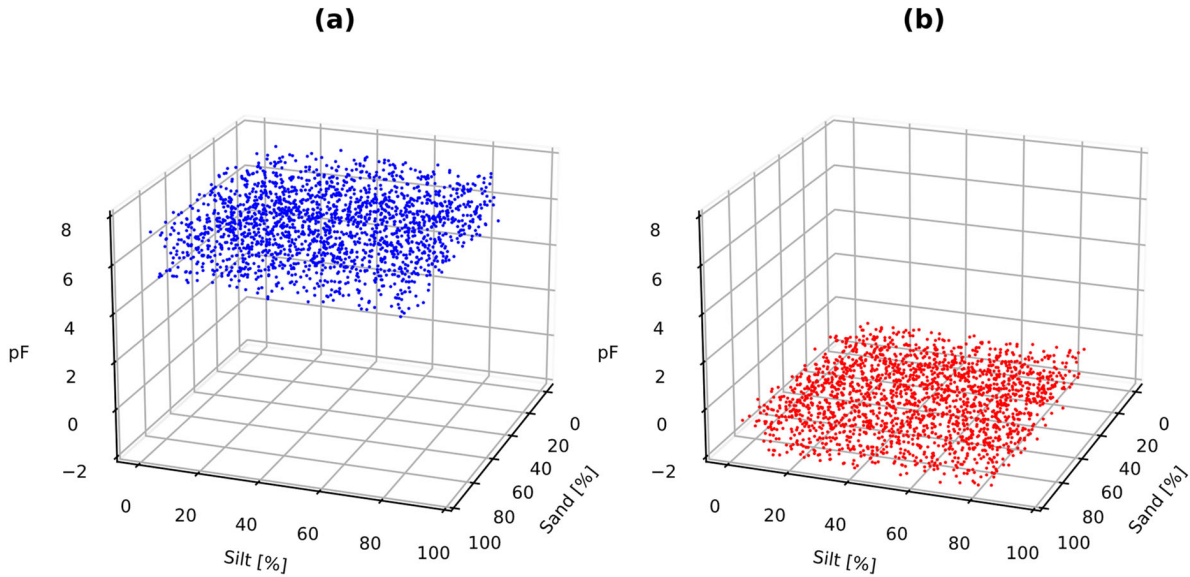

**Figure 2.** The two sets of residual (collocation) points used to enforce physical constraints: (a) zero capillary water for $pF > 5$ and (b) constant water content for $pF < -0.3$.

*2.2.7 Model training via automatic differentiation*

The hybrid model developed in this study involves several interconnected neural networks, each containing trainable parameters (Fig. 1). These networks are coupled in Eq. (8) through a physics-informed ansatz (i.e., a prior mathematical form assumed to guide the model). This coupling of various neural networks results in a highly nonlinear system whose parameters must be optimized to minimize the total loss function defined in Eq. (9).

Training such a hybrid system requires computing gradients of the loss function, Eq. (9), with respect to all trainable parameters. To enable efficient and accurate gradient computation, we leverage automatic differentiation (AD), a core feature in modern deep learning frameworks such as TensorFlow and PyTorch (Baydin et al., 2018). Automatic differentiation automatically constructs a computational graph during the model's forward pass and traces the sequence of mathematical operations from inputs to outputs. During backpropagation, reverse-mode AD traverses this graph from the output layer back to the inputs, systematically applying the chain rule to compute exact gradients with respect to every trainable parameter. This allows the model to be trained efficiently using standard gradient-based optimization algorithms, despite its architectural complexity and the presence of embedded physical constraints.

*2.2.8 Hyperparameter optimization*

The developed hybrid model consists of separate sub-network neural models, each designed for a specific component. Table 1 reports the inputs, architecture, and activation functions of each sub-network. All hidden layers across the sub-networks employed Exponential Linear Unit (ELU) activation functions.

The output of networks $NN_c$, $NN_s$, $NN_{dry}$, and $NN_o$ were implemented with sigmoid activation functions and were scaled appropriately to their respective physical ranges. Additionally, the networks $NN_s$, $NN_o$, and $NN_{dry}$, which estimate soil-dependent constants, shared a similar structure: two hidden layers with four units each and ELU activations.

We set $\lambda_1$ and $\lambda_2$ in Eq. (9) to 1 and 12.1, respectively, based on manual tuning. The parameters $\lambda_3$ and $\lambda_4$ were set to 10 and 5, respectively. These weights were tuned to ensure their corresponding constraints were satisfied without degrading the overall performance of the model. $S_1$ and $S_2$, which determine the number of residual points in sets 1 and 2, were both set to 2000.

The Adam optimizer (Kingma, 2014) was used for model training with an initial learning rate of 0.005. An adaptive learning rate strategy was applied, in which the learning rate was reduced by a factor of 0.8 if no improvement in the validation loss was observed, continuing down to a minimum of 0.0005.

To avoid overfitting, early stopping with a patience of 10 epochs was applied. Additionally, because the capillary sub-network ($NN_c$) directly influences the smoothness of the final model, we applied L2 regularization with an intensity of 0.15 to all layers of this sub-network. L2 regularization prevents overfitting by controlling the magnitude of large weights in the network (Ng, 2004). The model was entirely developed in Python and implemented using TensorFlow (Abadi et al., 2016).

## 2.3 Experimental data and train/validation/test splits

For training and evaluating the hybrid pedotransfer functions in this study, we used 482 undisturbed soil samples from the publicly available dataset of Hohenbrink et al. (2023), which include measurements of soil hydraulic properties for a wide range of texture types and organic carbon contents (Fig. 3 and Table 2). The 482 soils were extracted from the main dataset (consisting of 572 soil samples) with the condition that measurements of SWRC, soil textural fractions, bulk density, and organic carbon content, are available.

Particle size distribution was measured using wet sieving and sedimentation techniques. The size classes were categorized following the United States Department of Agriculture (USDA) classification system, which defines clay as particles smaller than 2 μm, silt as 2–50 μm, and sand as 50–2,000 µm (Hillel, 1982). The selected set of soil samples covers eleven soil textural classes of USDA system, making the dataset highly suitable for the data-driven approach in this study (Fig. 3).

Organic carbon content was measured by high-temperature combustion using an elemental analyzer. Bulk density was determined gravimetrically, by oven-drying the soil samples for at least 24 hours following evaporation experiments.

**Table 2.** Summary statistics of soil physical properties in the dataset

|      | Sand [%] | Silt [%] | Clay [%] | OC [%] | BD [gr cm$^{-3}$] |
|------|----------|----------|----------|--------|-------------------|
| Min  | 3.3      | 0.0      | 0.1      | 0.04   | 0.37              |
| Mean | 41.3     | 38.6     | 20.1     | 2.11   | 1.33              |
| Max  | 99.9     | 85.7     | 66.4     | 19.26  | 1.89              |
| Std  | 30.4     | 24.6     | 13.2     | 2.15   | 0.29              |

OC, organic carbon; BD, bulk density; Std, standard deviation

The dataset includes measurements of the soil water retention curve covering a broad range of matric heads. The wet and medium moisture ranges ($pF \leq 4.2$) were measured using the simplified evaporation method (Peters and Durner, 2008; Schindler, 1980), implemented via the HYPROP device (METER Group AG, Germany). This method captures the drying branch of the retention curve and provides high-resolution measurements. For the dry end ($pF > 4.2$), additional measurements were obtained using the dewpoint method (Campbell et al., 2007; Kirste et al., 2019) via the WP4C device (METER Group, Inc., USA). Due to the large imbalance between the number of dry-end and wet-end data points, we uniformly resampled the wet-end measurements to 15 points. For more detailed information about the measurement details and device specifications, readers are referred to Hohenbrink et al. (2023).

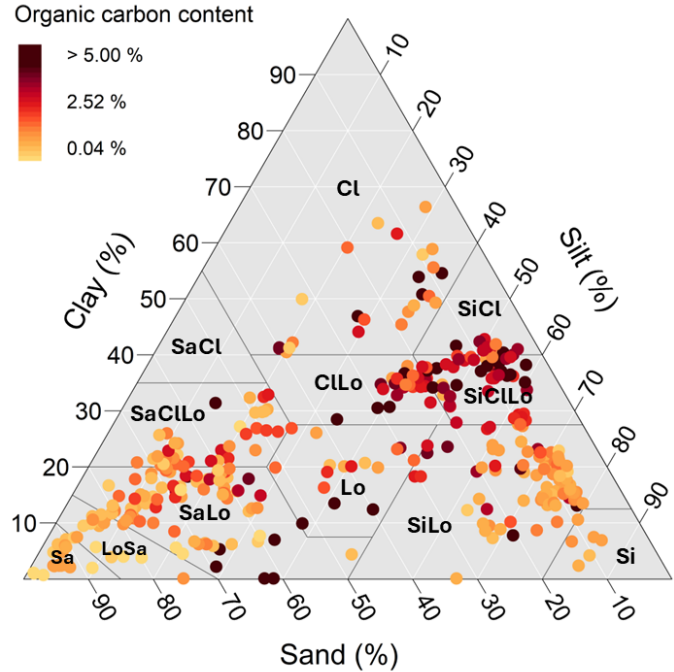

**Figure 3.** The textural distribution of 482 soil samples according to the USDA (United States Department of Agriculture) soil triangle. Point colors indicate organic carbon content (%) of each soil sample. The abbreviations Sa, Si, and Cl stand for sand, silt, and clay, respectively.

To train the model with the customized loss function in Eq. (9), which includes separate terms for the wet and dry ends of the SWRC, we partitioned the data for each end into training and test sets separately to achieve a more balanced split. Seventy percent of the samples from each end were used for training and validation, while the remaining 30% served as a hold-out test set. This stratified sampling ensured that each subset, training, validation, and testing, represented the full range of wet and dry conditions. The dataset comprised 8394 $pF - \theta$ observations after resampling, of which 5916 $pF - \theta$ pairs (339 samples) were used for training and validation, and 2478 $pF - \theta$ pairs (143 samples) for testing.

To avoid data leakage and account for the strong correlation among water content measurements within individual soil samples, we partitioned the dataset at the soil sample level rather than at the level of individual $pF - \theta$ pairs. This ensured that all $pF - \theta$ observations from a given sample were assigned entirely to either the training or testing set. Additionally, we ensured that samples from the same soil profile (i.e., location) were exclusively included in either the training or the testing set.

*2.4 Evaluation criteria*

Model performance was evaluated based on three metrics: root mean square error (RMSE), coefficient of determination (R-squared), and the ratio of the interquartile range to RMSE (RPIQ), all calculated using the predicted and observed volumetric water contents:

$$RMSE = \sqrt{\frac{1}{N}\sum_{i=1}^{N}(\hat{\theta}^{(i)} - \theta^{(i)})^2} \tag{11}$$

$$R^2 = 1 - \frac{\sum_{i=1}^{N}(\hat{\theta}^{(i)} - \theta^{(i)})^2}{\sum_{i=1}^{N}(\theta^{(i)} - \bar{\theta})^2} \tag{12}$$

$$RPIQ = \frac{Q_{75} - Q_{25}}{RMSE} \tag{13}$$

where $N$ is the total number of measured points, $\bar{\theta}$ is the mean of the measured water contents, and $Q_{75}$ and $Q_{25}$ correspond to the 75th and 25th percentiles of the measured water contents, respectively. The RPIQ metric offers a scale-independent metric by comparing the RMSE with the variability of the data.

## 3 Results and Discussion

### *3.1 Physics-informed neural network performance*

Figure 4 shows the predicted shape of the SWRC for twelve soil samples with different texture classes according to the USDA classification system. The model is trained by optimizing the loss function (Eq. (9)) over the entire training set as a whole, rather than fitting it sample by sample. Unlike parametric PTFs, which use predefined analytical forms for the SWRC, our hybrid model learns the curve shape directly from the data. Once trained, the model can predict the entire continuous SWRC. To achieve this, for any given soil with fixed physical properties (i.e., sand, silt, clay, OC, and BD), we vary the $pF$ over a specified range to generate the continuous curves shown in Fig. 4.

The discovered shapes of the SWRCs are smooth, analytically differentiable, and therefore suitable for simulation of soil water flow (i.e., Richardson-Richards equation). These curves exhibit a sigmoidal shape in the wet range, similar to traditional parametric models (van Genuchten, 1980; Kosugi, 1994), and transition to a linear form at lower water contents, consistent with the Campbell-Shiozawa model behavior assumed at the dry end. Notably, the transition between the neural network-predicted region and the analytically modeled region governed by the Campbell-Shiozawa model is seamless, with no noticeable discontinuities or abrupt changes, resulting in a smooth, continuous curve. Furthermore, at the wet end, the curves remain invariant with respect to $pF$ for values below -0.3. At the dry end, the range of $pF$ at zero water content, $pF_{dry}$, for all curves remains between 6.2 and 7.6, ensuring the satisfaction of the physical constraints imposed at both ends during training.

The overall performance of the hybrid PTFs on both the train and test sets is depicted in Fig. 5. As shown, the model demonstrates a close performance on both the train and test sets, indicating the generalization capability of the model. This close performance between training and testing sets is particularly important in developing continuous, non-parametric and semi-parametric PTFs, as even small degrees of overfitting can distort the predicted curve, making it physically unrealistic.

The model achieved an RMSE of 0.049 m³ m⁻³ on the test set, which is reasonable given the diversity of soils represented in the dataset, including eleven USDA texture classes, and samples with high organic carbon content and very low bulk density (Table 2). The extent of variations in soil properties is also reflected in Fig. 5, where measured water contents reach values as high as 0.8 $m^3$ $m^{-3}$. This diversity, along with dataset size, variations in soil properties, and measurement quality, are key factors influencing the performance of PTFs. The performance of the PTF developed aligns well with both continuous parametric models (dos Santos Pereira et al., 2025) and continuous non-parametric models trained on HYPROP system measurements (Haghverdi et al., 2018). However, due to differences in the datasets used and input predictors across studies, direct comparisons are challenging.

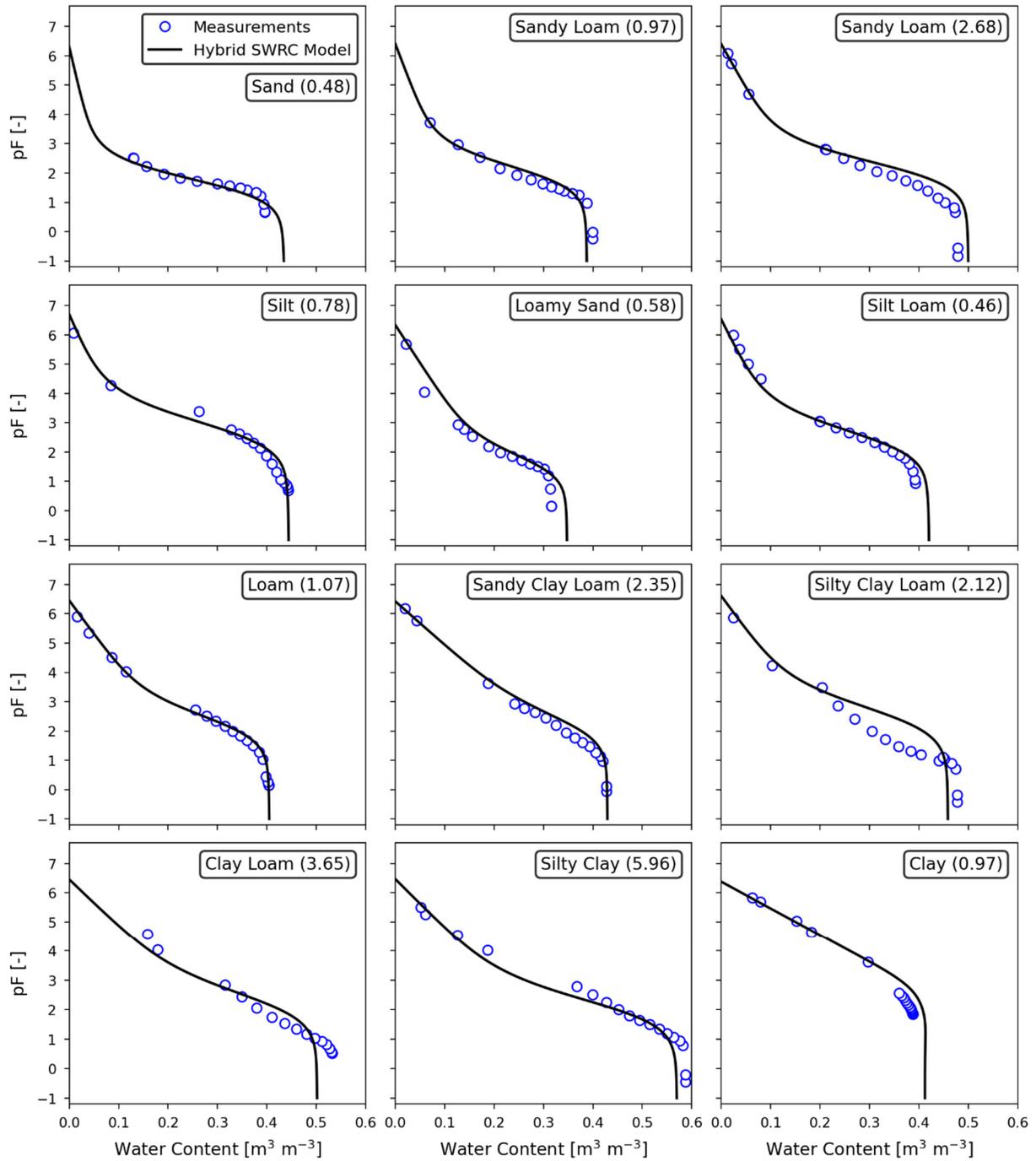

**Figure 4.** Predicted soil water retention curves from the developed hybrid pedotransfer functions for twelve soils representing different USDA textural classes. Values in parentheses represent the organic carbon content (OC) in percentage.

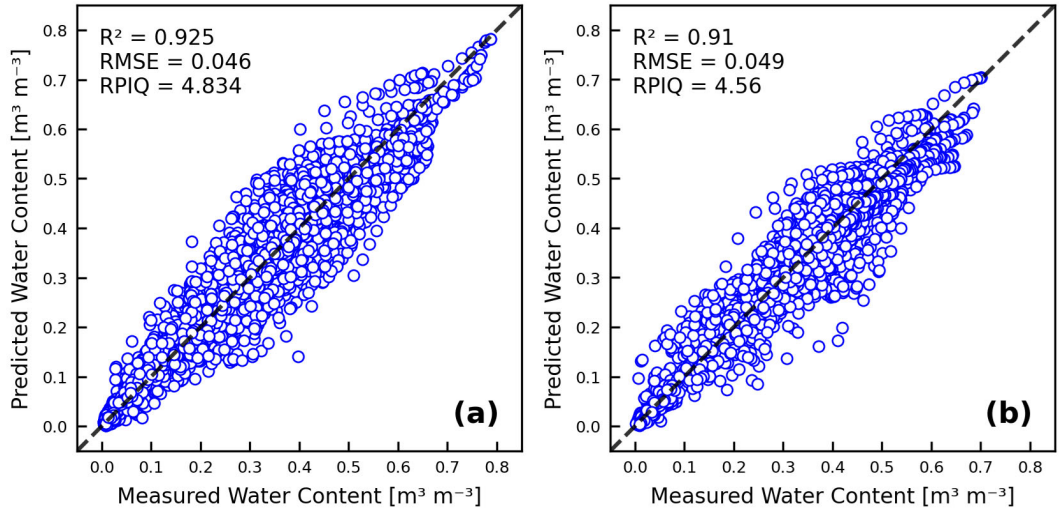

**Figure 5.** Overall performance of the hybrid pedotransfer functions on the (a) train and (b) test sets.

The discovered partitioning of capillary and adsorbed film components of the SWRC for the soil samples shown in Fig. 4 is depicted in Fig. 6. These curves are obtained by plotting the first and second terms on the right-hand side of Eq. (8). The curves shown in Fig. 6 should be considered as follows: at each $pF$, the total water content (black line) is the sum of the capillary and adsorbed film water contents.

This data-driven partitioning aligns remarkably well with the physics-based models in the literature, such as that of Or and Tuller (1999) model, which was developed by incorporating detailed interfacial physics within an angular pore geometry. Specifically, the capillary component dominates under saturated conditions for all soils. This corresponds to the point where the liquid–vapor interfacial area is effectively zero. As $pF$ increases (corresponding to more suction in soil), pores of varying sizes begin to drain, and this process starts with larger pores. For each pore size, there exists a critical $pF$, often referred to as the "onset of drainage", at which air starts to invade the pore.

As drainage progresses, water films begin to form along the surfaces of the partially emptied pores. With further increase of $pF$, smaller pores also undergo drainage, leading to a gradual decrease in the capillary component and a concurrent increase in the contribution of the film component. This reflects the physical process by which surface area becomes increasingly available for film water as the capillary water recedes to the pores corners (Or and Tuller, 1999).

At a $pF$ between 2.5 and 4 for all soils, the film component reaches a peak. Beyond this peak, toward the dry region, both capillary and adsorbed film contributions decline; this is the point where even the adsorbed water films begin to thin. This decline continues at very high $pF$ values and eventually the adsorbed water converges to the Campbell-Shiozawa linear model for the dry end, with the extent of this linear region being dependent on soil texture (Fig. 6). Note that, consistent with physical constraints imposed, the water content for $pF > 5$ remains entirely in adsorbed form for all soils.

Another important point in Fig. 6 is the crossover between capillary and adsorbed component curves, which determines the boundary between capillary and adsorbed dominant regions. This crossover point is highly dependent on soil texture, and as seen in Fig. 7, with an increase in clay content, the water content at this point increases. This is because the fraction of finer particles (i.e., silt and clay) provides a greater specific surface area, which supports the formation and retention of more extensive water films along particle surfaces (Norouzi et al., 2022, 2023).

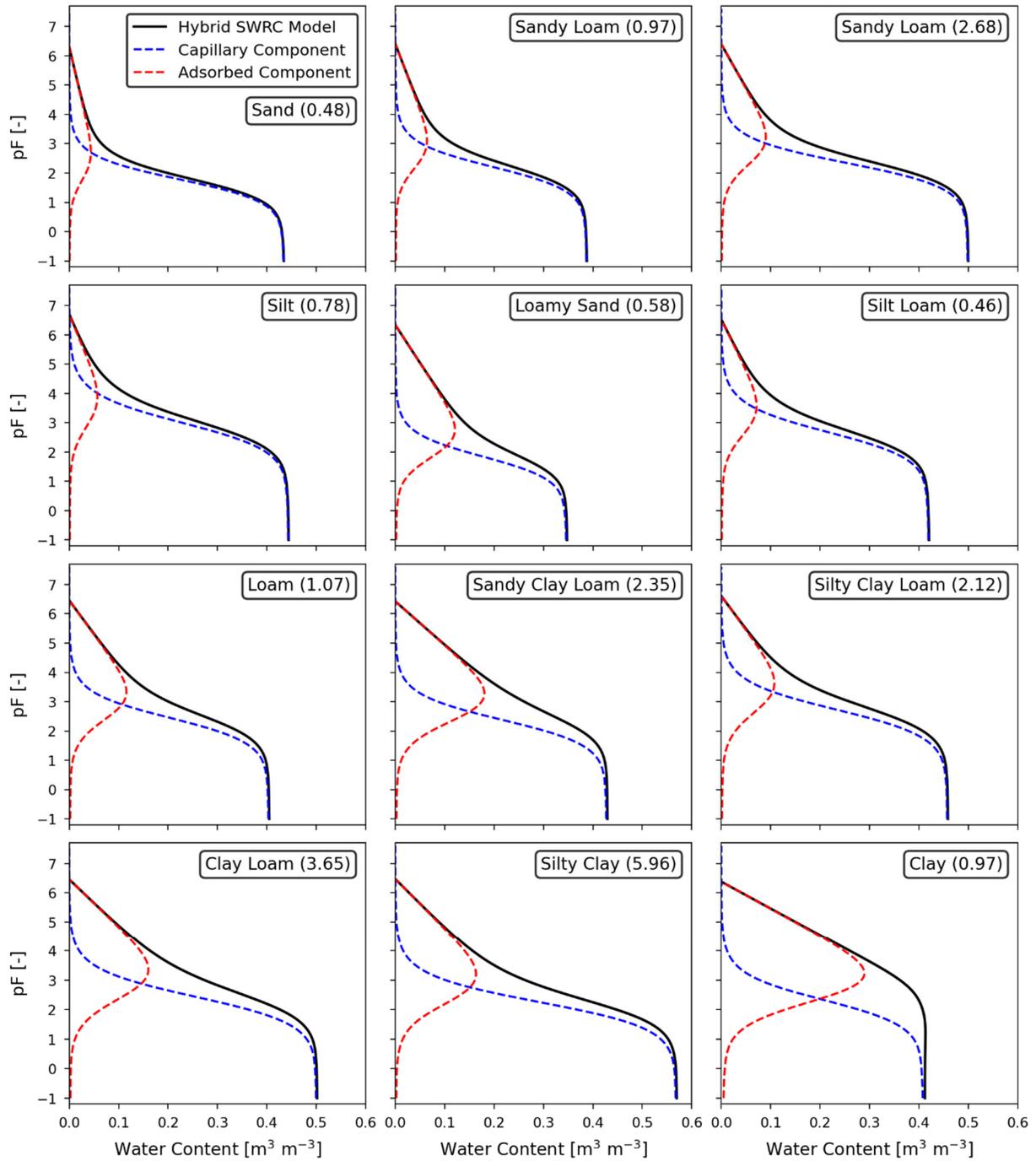

**Figure 6**. Data-driven discovery of capillary and adsorbed film components by the hybrid model for the same soil samples shown in Fig. 4. Values in parentheses represent the organic carbon content (OC) in percentage.

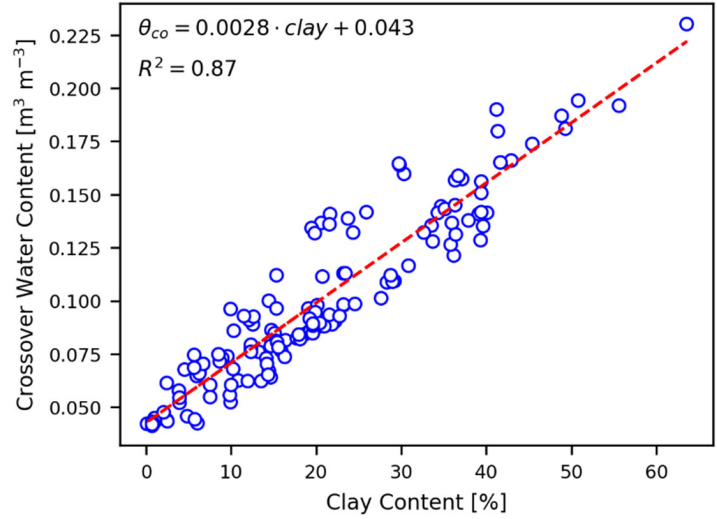

**Figure 7.** Water content at the crossover point between capillary and adsorbed film components, denoted as $\theta_{co}$, versus clay content for all soils in the test set.

An important advantage of DHM is the ability to replace unknown or poorly defined components with neural networks, which serve as universal function approximators. In our model, the transition function $f(S_c)$ in Eq. (5), whose analytical form was not known in advance, was learned directly from data. This function governs how the Campbell-Shiozawa model should be modified at $pF$ values lower than a certain threshold, where capillary and adsorbed film water may coexist. This learned function is shown in Fig. 8. As illustrated, $f(S_c)$ exhibits a decreasing trend, which is expected: as capillary water increases, the contribution of the Campbell-Shiozawa model to total water content should diminish.

The transition function was designed with a hard constraint to satisfy $f(S_c = 0) = 1$, ensuring full reliance on the Campbell-Shiozawa model when the capillary saturation is zero (Fig. 8). Interestingly, although not explicitly constrained to do so, the learned function also satisfies $f(S_c = 1) = 0$. This behavior implies that at saturation ($S_c = 1$), the entire water content is attributed to capillary water, with no contribution from the adsorbed film component (Fig. 6). Additionally, the transition between the two endpoints ($S_c = 0$ and $S_c = 1$) is distinctly nonlinear, deviating from the commonly assumed linear transition ($1 - S_c$) used in previous studies (Fayer and Simmons, 1995; Lebeau and Konrad, 2010).

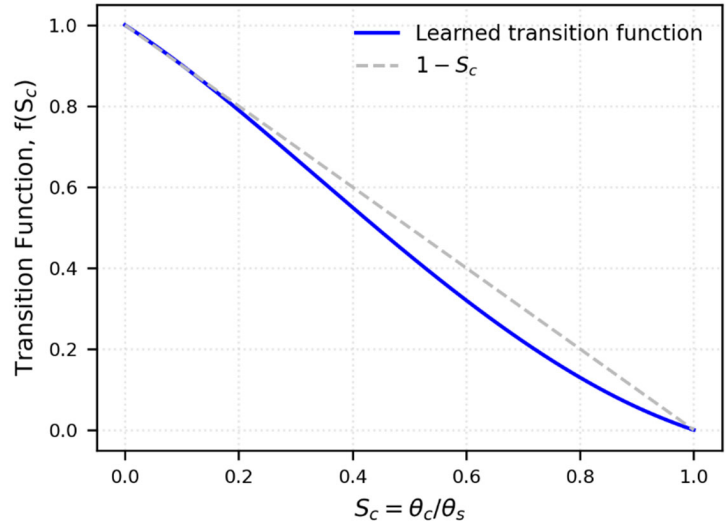

**Figure 8.** Learned transition function, $f(S_c)$, as defined in Eq. (5). The reference curve $1 - S_c$, commonly assumed in previous studies, is shown in gray to illustrate the deviation of the learned function from linear behavior.

*3.2 Performance of the sub-networks for soil constants*

One of the key advantages of DHM is its ability to train multiple internal models simultaneously within a unified framework (Shen et al., 2023). In our hybrid PTF, we implemented three dedicated sub-networks (i.e., sub-PTFs) to estimate key soil parameters, namely, $\theta_s$, $pF_{dry}$, and $\theta_o$, directly from basic soil physical properties (see Eq. (8) and Table 1). Each of these sub-networks maps soil properties to a physical constant. After training, each of these networks could be used as a standalone PTF. It should be noted that these sub-networks were not trained with separate target data; instead, they adjusted their parameters as part of the joint training of the full hybrid model (Fig. 1).

Figure 9 illustrates the predictions of each sub-network, plotted against one of the representative input variables, to ensure that these sub-networks have learned meaningful physically or empirically known relationships rather than overfitting the training data.

The sub-network $NN_s$ predicts $\theta_s$. The predictions of this sub-network exhibit a clear inverse relationship with bulk density in its input layer, consistent with the physical understanding that higher bulk density typically corresponds to lower total porosity and thus lower water content at saturation (Fig. 9a). Interestingly, for high bulk density values, the predictions align well with the standard equation used to calculate porosity from known bulk density, assuming a particle density of 2.65 g/cm³. However, for soils with low bulk density (organic soils), the predictions deviate from this relationship, demonstrating the inadequacy of the 2.65 g/cm³ assumption for such soils (Marakkala Manage et al., 2019).

In Fig. (9b) the predictions of $NN_s$ are plotted against OC in the input layer. As expected, saturated water content increases with increasing OC. Organic particles have lower intrinsic density with more irregular, often fibrous structures than mineral particles, which leads to looser packing and greater total pore volume. Consequently, soils with higher organic carbon content tend to exhibit higher porosity and lower bulk density, consistent with observations in analytical modeling studies (compare for example with the trend observed in Figure 4 of Robinson et al. (2022)).

Similarly, the sub-network $NN_o$ (Fig. 9c), which predicts $\theta_o$ of Campbell-Shiozawa model and determines the slope of the linear region at the dry end, shows a positive correlation with clay content. As noted by Campbell and Shiozawa (1992), this free parameter is highly influenced by the amount of soil clay content, which determines the specific area of soil. Higher $\theta_o$ values are indicative of finer-textured soils, which retain more water across a wide range of matric heads.

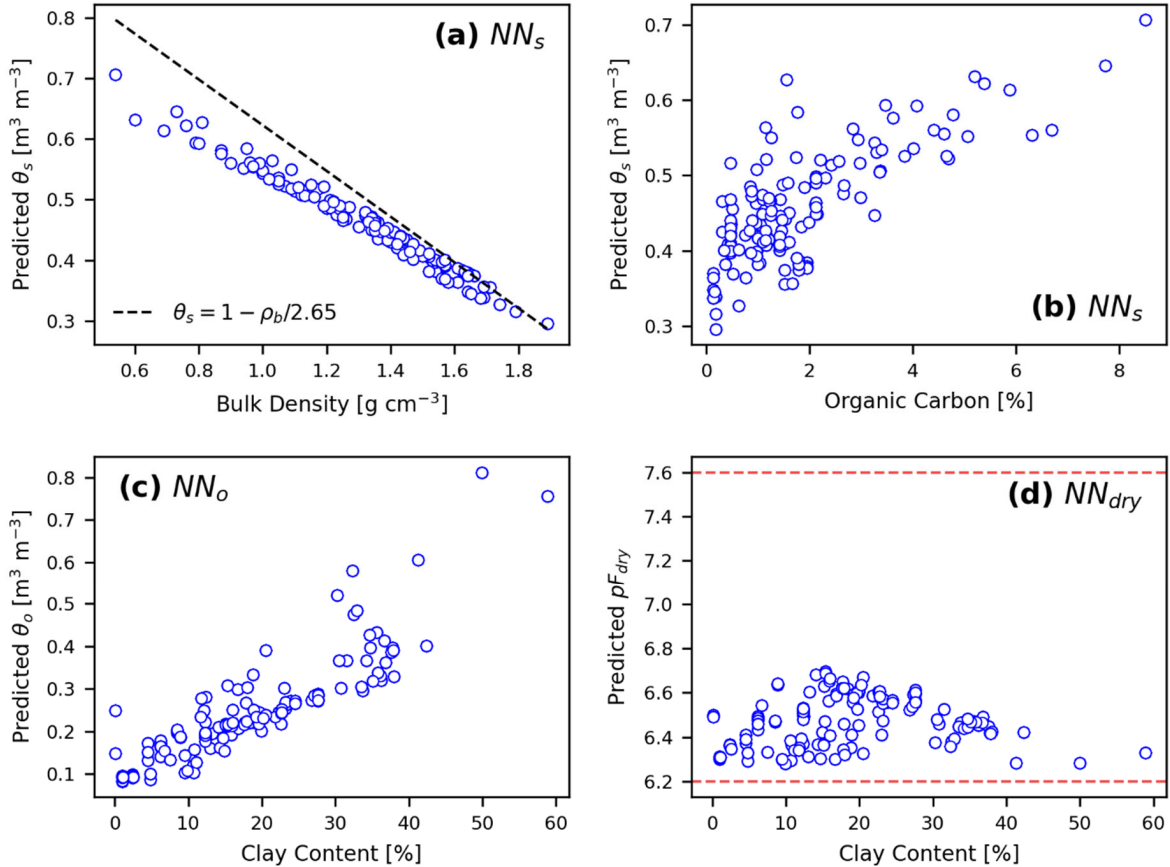

**Figure 9.** Predicted soil-specific parameters by the interior sub-networks of the hybrid PTF model, plotted against representative input variables. The parameter $\theta_s$ represents saturated water content and $\theta_o$ and $pF_{dry}$ are the fitting parameters that determine the slope and intercept of the empirical model at the dry end [Eq. (2)].

In contrast to the other two sub-networks, the sub-network predicting $pF_{dry}$, denoted as $NN_{dry}$, does not show any clear correlation with the input variables. This is consistent with findings by Lu and Khorshidi (2015) and Karup et al. (2017), who showed that $pF_{dry}$ is more dependent on clay mineralogy than soil OC or clay content. Although no strong correlation is observed, all predictions from this sub-network remain within the imposed limits ($pF = 6.2$ to $7.6$), which were enforced as a hard constraint by scaling the network's output to the target range (Fig. 9d).

These results collectively demonstrate that the interior sub-networks not only remained within the specified physical bounds but also captured relationships that are consistent with established physical and empirical understanding.

## 4 Conclusions and outlook

We have introduced differentiable hybrid modeling (DHM) framework that integrates mechanistic understanding with data-driven learning in a unified framework. In this method the unknown or less understood parts of a system can be learned from data by embedding neural networks while the known physical processes are explicitly preserved.

As a proof of concept, we applied the DHM framework to the challenging problem of partitioning the soil water retention curve (SWRC) into capillary and adsorbed components. At the dry end, where there is general agreement on the dominance of the adsorbed water component, we used an analytical model, while for higher saturations with more complex processes, we embedded physics-informed neural networks. Our new generation of SWRC models, which we refer to as "semi-parametric models", successfully learned the overall shape of the SWRC as well as its adsorbed and capillary component. Notably, the hybrid model learned pore-scale features without relying on explicit geometrical assumptions about soil pore shape or its distribution.

Our model demonstrates a new perspective on the use of data in soil physics. We used the same inputs and outputs as conventional pedotransfer functions, translating basic soil physical properties into a soil water retention curve, but our main objective goes far beyond simple prediction. During this mapping from inputs to outputs, the hybrid model learns multiple intermediate processes implicitly, without requiring explicit data for them. Importantly, these learned internal relationships produced physically meaningful results and as observed in the case of transition function, the model discovered a nonlinear function that challenges the linear assumptions invoked in previous studies. This represents a fundamental shift in how pedotransfer functions can be used, not merely to predict, but to discover the underlying physical relationships encoded in data.

A great advantage of the DHM framework is its flexibility in design. As our understanding of soil physical processes advances, we can incorporate more soil physics knowledge into the hybrid model structure. For example, in this study we used the Campbell-Shiozawa analytical model to represent the dry end, owing to its better fit to the approximately linear behavior observed in that region. Future work could examine the use of more physically based dry-end formulations within the same DHM workflow (Tuller and Or, 2005b; Smagin, 2025). From another perspective, the internal components trained in our model could later be used as standalone predictive models for specific soil properties or integrated into other modeling frameworks.

Given the increasing availability of large soil datasets, we believe DHM and its capability for end-to-end training of several internal components (i.e., sub-models) within a single optimization process, provides a promising tool that can be leveraged for modeling fundamental physical processes where partial knowledge of the underlying mechanisms has led to over-simplifying assumptions and biased predictions.

**Acknowledgments**

The authors gratefully acknowledge the late Professor Markus Tuller, whose mentorship and insightful collaborations greatly influenced the ideas and results presented in this work.

This study is funded by the European Union Horizon Europe research and innovation program under grant agreement No. 101086179 (AI4SoilHealth). Views and opinions expressed are, however, those of the author(s) only and do not necessarily reflect those of the European Union or the Research Executive Agency (REA). Neither the European Union nor the granting authority can be held responsible for them.